\documentstyle[aps,pre,12pt]{revtex}
\newcommand{\perpp}[0]{{\perp}}
\newcommand{\dnb}[0]{\delta{\vec n}}
\newcommand{\kb}[0]{k_{\rm B}}
\newcommand{\half}[0]{\frac{1}{2}}
\newcommand{\nb}[0]{{\bf n}}
\newcommand{\bo}[1]{{\cal O}(#1)}
\newcommand{\nablab}[1]{\setbox0=\hbox{$\nabla$}%
     \kern-.010em\copy0\kern-\wd0
     \kern.025em\copy0\kern-\wd0
     \kern-.020em\raise.0200em\box0 }
\begin{document}
\preprint{\hbox{IASSNS-HEP-93/28, CMT-HU-N02}}
\title{Rotational Invariance and the Theory of
Directed Polymer Nematics}
\author{Randall D. Kamien}
\address{School of Natural Sciences, Institute for Advanced Study,
Princeton, NJ 08540}
\author{Pierre Le Doussal\footnote{On leave from Laboratoire de Physique
Theorique de l'Ecole Normale Sup{\'e}rieure Paris,
Ecole Normale Sup{\'e}rieure, 24 rue Lhomond, Paris 75231 Cedex 05, France.
(Laboratoire Propre du CNRS).}}
\address{Serin Physics Laboratory, Rutgers University, Piscataway,
NJ 08855}
\author{David R. Nelson}
\address{Lyman Laboratory of Physics, Harvard University,
Cambridge, MA 02138}
\date{8 June 1993}
\maketitle
\begin{abstract}
The consequences of rotational invariance in a recent theory of fluctuations in
dilute polymer
nematics are explored.  A correct rotationally invariant free energy insures
that anomalous
couplings are not generated in a one-loop renormalization group calculation.
\end{abstract}
%PACS numbers: 36.20.Ey, 05.40.+j
\section{Introduction and Summary}

In this addendum we refine an earlier theory of polymer nematics
\cite{KLN}. In that paper a renormalization group calculation was performed in
the dilute
polymer limit which
reproduced the anomalous, logarithmic wandering predicted by de Gennes
\cite{Ra}.

We analyzed the model defined by the grand-canonical partition function:
\begin{equation}
{\cal Z}_{gr} = \int {\cal D}\psi{\cal D}\psi^*{\cal D}\dnb\,\exp
\left\{-S[\psi,\psi^*,\dnb]\right\} ,
\label{epart}
\end{equation}
where the action was broken into three parts,
\begin{eqnarray}
S[\psi,\psi^*,\dnb]&&=\int d^d\!r\int dz\,\left[\psi^*\left(\partial_z
-D\nabla_\perpp^2 -\bar\mu\right)\psi + v\vert\psi\vert^4\right]\nonumber\\
&&\qquad+\;{\lambda\over 2}\int d^d\!r\int
dz\,\dnb\cdot\left(\psi^*\nabla_\perpp\psi
-\psi\nabla_\perpp\psi^*\right)\nonumber\\
&&\qquad+\; F_n[\dnb]/\kb T,
\label{threeparts}
\end{eqnarray}
and
\begin{equation}
F_n = {1\over 2}\int d^2\!r_{\perpp}\int dz
\,\left[K_1(\nabla_{\perpp}\!\cdot\!
\dnb)^2 + K_2(\nabla_{\perpp}\!\times\!\dnb)^2 + K_3(\partial_z
\dnb)^2\right],
\label{eIi}
\end{equation}
where $\psi$ is the ``boson'' order parameter describing the directed polymers
and $\dnb$ is
a $d$-dimensional vector which
describes director fluctuations in the nematic matrix.

We performed a renormalization group calculation near $\mu=0$ and found
logarithmic
corrections
to mean field theory in the critical dimension $d_c=d+1\equiv 2+1$.  We show
here that an
additional
$\vert\psi\vert^2\dnb^2$ term, which appears to be generated at one-loop order,
is
canceled by additional coupling required by rotational invariance.
We derived Eq. (\ref{epart}) starting with a rotationally
invariant theory by choosing a broken symmetry direction for the nematic (which
we take to be the $z$-axis), and
expanding to quadratic order in fields.  If we consider the nematic as an
external,
non-fluctuating field, then Eq. (\ref{epart}) should be invariant under the
following
transformation:
\begin{eqnarray}
\dnb'({\vec r}\; ',z') && = \dnb(\vec r,z) + \vec h\nonumber\\
\psi'({\vec r}\; ',z') && = \psi(\vec r,z)\nonumber\\
{\vec r}\; ' &&= \vec r + \lambda\vec h z\nonumber\\
z'&&=z
\label{eol}
\end{eqnarray}
In the limit of small $\vec h$, this affine change of variables is equivalent
to
rotating the system by an amount $\vec h$.
With this change of co\" ordinates, $\partial_{z'}=\partial_z-\lambda\vec
h\cdot\nabla_\perpp$,
$\nabla_\perpp'=\nabla_\perpp$ and $d^d\!r'dz'=d^d\!rdz$, and the part of the
action $S_1$ which depends on $\psi$ becomes
$S_1$
\begin{eqnarray}
S_1'&[\psi',\dnb']  \nonumber\\
&&=\int d^d\!r'dz'\,\left[{\psi^*}'\left(\partial_z' - D(\nabla'_\perpp)^2
-\bar\mu\right)\psi'
+v\vert\psi'\vert^4 +{\lambda\over
2}\dnb'\cdot\left({\psi^*}'\nabla'_\perpp\psi'-\psi'
\nabla'_\perpp{\psi^*}'\right)\right]\nonumber\\
&&= S_1[\psi,\dnb] + \int d^d\!rdz\,\left[\psi^*\left(-\lambda\vec
h\cdot\nabla_\perpp
\right)\psi  +{\lambda\over 2}\vec h
\cdot\left(\psi^*\nabla_\perpp\psi-\psi\nabla_\perpp\psi^*
\right)\right]\nonumber\\
&&= S_1[\psi,\dnb]+\hbox{surface terms}
\label{ess}
\end{eqnarray}
Under this symmetry, a term of the form $\vert\psi\vert^2\dnb^2$ would be
forbidden,
as a shift in $\dnb$ would generate additional terms.

However, when the Frank free energy is included in the full action $S$,
it is not invariant under
Eq. (\ref{eol}).  In particular
\begin{eqnarray}
F'_n[\dnb'] &&= \half\int d^d\!r'dz'\,\left[
K_1(\nabla_\perpp'\cdot\dnb')^2 + K_2(\nabla_\perpp'\times\dnb')^2 +
K_3(\partial_{z'}\dnb')^2
\right]\nonumber\\
&&=\half\int d^d\!rdz\,\left[
K_1(\nabla_\perpp\cdot\dnb)^2 + K_2(\nabla_\perpp\times\dnb)^2 +
K_3(\partial_z\dnb-\lambda
\vec h\cdot\nabla_\perpp\dnb)^2
\right]
\label{efre}
\end{eqnarray}
Because the Frank free energy is not invariant, the new relevant operator is
generated
(see figure 1).
In the boson analogy, the symmetry above corresponds to the Galilean invariance
of
the original polymer action (a fully rotationally invariant polymer theory
would correspond to a relativistically invariant boson theory).
Consider the fully rotationally invariant Frank energy
\begin{equation}
F_n[{\bf n}] = \half\int d^2\!rdz\,\left[K_1(\nablab\cdot\nb)^2 +
K_2\left(\nb\cdot
(\nablab\times\nb)\right)^2 +
K_3\left(\nb\times(\nablab\times\nb)\right)^2\right]
\label{efull}
\end{equation}
where boldface represents three-dimensional vectors and $\nb$ is
a unit vector.  We must expand Eq. (\ref{efull}) in powers of $\dnb$ so that it
too is
invariant under Eq. (\ref{eol}).
Upon expanding Eq. (\ref{efull}) we arrive at
\begin{equation}
F_n[\dnb] \approx \half\int d^2\!rdz\,\left[K_1(\nabla_\perpp\cdot\dnb)^2
+K_2(\nabla_\perpp\times\dnb)^2
+K_3\left(\partial_z\dnb + (\dnb\cdot\nabla_\perpp)\dnb\right)^2\right]
\label{eorderc}
\end{equation}
The new form for the bend term is not unlike terms required by
rotational invariance in smectics \cite{GP} and polymer nematics \cite{TON}.
In fact, in \cite{TON}, the
corrections due to rotational invariance were also irrelevant at one-loop
and converged in $2+1$ dimensions.
We had
originally added a factor of $\lambda$ in the interaction
between $\dnb$ and $\psi$ to organize perturbation theory.  This
amounted to rescaling $\dnb$ and the Frank constants $K_i$.  Doing this
with the new bend term gives
\begin{equation}
F_{\hbox{bend}} = \half\int d^d\!rdz\,K_3
\left(\partial_z\dnb +\lambda(\dnb\cdot\nabla_\perpp)\dnb\right)^2
\label{ebend}
\end{equation}
We have repeated
the renormalization group calculation and find, to one-loop order, that the
additional
terms in Eq. (\ref{efull}) are irrelevant, and again, the results in \cite{KLN}
are, again, correct.

Though the symmetry prevents $\vert\psi\vert^2\dnb^2$ from appearing in the
action, in
perturbation theory this amounts to a delicate cancellation among graphs (see
figure 2).
Checking this explicitly, we find that these graphs do indeed cancel, and our
original results are unchanged.

\acknowledgements

RDK would like to thank M.~Goulian for enlightening discussions.
RDK was supported in part by the National Science Foundation, through Grant
No.~PHY92--45317, and through the Ambrose Monell Foundation.  PLD was
supported by the National Science Foundation, through Grant No.~4-20508.
DRN was supported by
the National Science Foundation, through Grant No.~DMR91--15491, and through
the Harvard
Materials Research Laboratory.

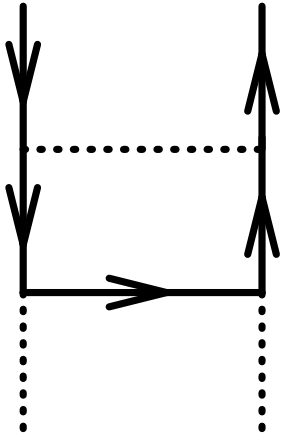
\begin{figure}
\caption{One-loop graph in original model which generates a term
$\vert\psi\vert^2\dnb^2$.  It is finite in $2+1$ dimensions.}
\label{figi}
\end{figure}
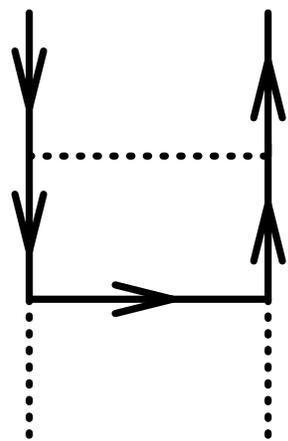
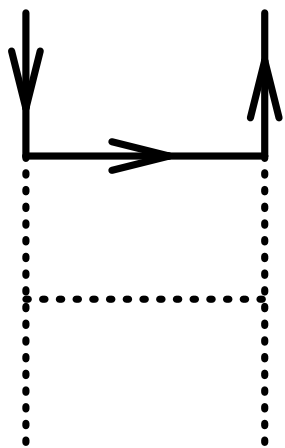
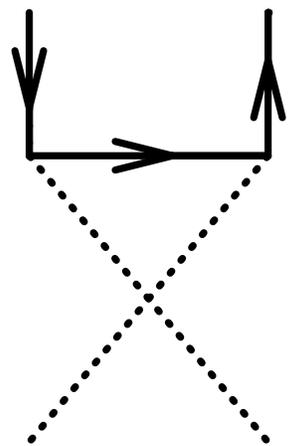
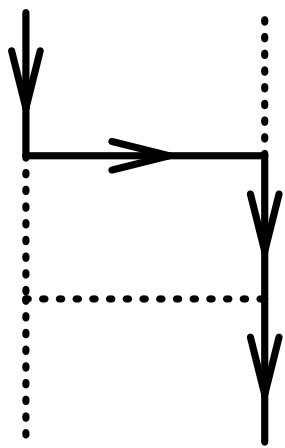
\begin{figure}
\caption{Graphs which all contribute to $\vert\psi\vert^2\dnb^2$ in
rotationally invariant theory.  Note that they are all $\bo{\lambda^4}$.
These graphs cancel among each other
to prevent such a term.}
\label{figii}
\end{figure}

\end{document}